\documentclass[onecolumn,epjc3]{svjour3} 
 
\usepackage[T1]{fontenc}
\usepackage{epsfig,amsmath,amssymb}
\usepackage{graphicx}
\usepackage{caption}
\usepackage{subcaption}
\usepackage{xspace}
\usepackage[listings]{tcolorbox}
\usepackage{ulem}
\usepackage{slashed}
\usepackage{cite}
\usepackage{xcolor}
\usepackage{units}
\usepackage{nicefrac}
\usepackage{textcomp}
\usepackage[utf8]{inputenc}
\usepackage{feynmp}
\usepackage[colorlinks,citecolor=blue,urlcolor=blue,linkcolor=blue]{hyperref}
\usepackage[capitalise]{cleveref}

\crefname{section}{Sec.}{Secs.}
\crefname{table}{Tab.}{Tabs.}
\crefname{figure}{Fig.}{Figs.}
\crefname{equation}{Eq.}{Eqs.}
\crefname{appendix}{Appendix\ }{Appendix\ }

\usepackage[framemethod=TikZ]{mdframed}
\usepackage{tikz} 
\usetikzlibrary{chains,shapes,fit,calc,snakes}
\usetikzlibrary{positioning,arrows}
\journalname{Eur. Phys. J. C}

\definecolor{maroon}{cmyk}{0, 0.87, 0.68, 0.32}
\definecolor{halfgray}{gray}{0.55}
\definecolor{slha_frame}{RGB}{207, 207, 207}
\definecolor{slha_bg}{RGB}{247, 247, 247}
\definecolor{slha_red}{RGB}{186, 33, 33}
\definecolor{slha_green}{RGB}{0, 128, 0}
\definecolor{slha_cyan}{RGB}{64, 128, 128}
\definecolor{slha_purple}{RGB}{170, 34, 255}

\definecolor{mathematica_frame}{RGB}{207, 207, 207}
\definecolor{mathematica_bg}{RGB}{247, 247, 247}
\definecolor{mathematica_red}{RGB}{186, 33, 33}
\definecolor{mathematica_green}{RGB}{0, 128, 0}
\definecolor{mathematica_cyan}{RGB}{64, 128, 128}
\definecolor{mathematica_purple}{RGB}{170, 34, 255}

\lstnewenvironment{MIN}[1][]{%
\mdframed[roundcorner=5pt,backgroundcolor=mathematica_bg]
  \renewcommand{\thelstnumber}{In[\arabic{lstnumber}]}
  \lstset{language=MathIn,numbers=left,basicstyle=\ttfamily,#1,escapeinside=||}%
}{%
\endmdframed
}

\lstnewenvironment{MOUT}[1][]{%
\mdframed[roundcorner=5pt,backgroundcolor=mathematica_bg]
  \renewcommand{\thelstnumber}{Out[\arabic{lstnumber}]}
  \lstset{language=MathOut,numbers=left,basicstyle=\ttfamily,#1}%
}{%
\endmdframed
}

\definecolor{maroon}{cmyk}{0, 0.87, 0.68, 0.32}
\definecolor{halfgray}{gray}{0.55}
\definecolor{ipython_frame}{RGB}{207, 207, 207}
\definecolor{ipython_bg}{RGB}{247, 247, 247}
\definecolor{ipython_red}{RGB}{186, 33, 33}
\definecolor{ipython_green}{RGB}{0, 128, 0}
\definecolor{ipython_cyan}{RGB}{64, 128, 128}
\definecolor{ipython_purple}{RGB}{170, 34, 255}

\lstdefinelanguage{iPython}{
    morekeywords={access,and,break,class,continue,def,del,elif,else,except,exec,finally,for,from,global,if,import,in,is,lambda,not,or,pass,print,raise,return,try,while},%
    morekeywords=[2]{abs,all,any,basestring,bin,bool,bytearray,callable,chr,classmethod,cmp,compile,complex,delattr,dict,dir,divmod,enumerate,eval,execfile,file,filter,float,format,frozenset,getattr,globals,hasattr,hash,help,hex,id,input,int,isinstance,issubclass,iter,len,list,locals,long,map,max,memoryview,min,next,object,oct,open,ord,pow,property,range,raw_input,reduce,reload,repr,reversed,round,set,setattr,slice,sorted,staticmethod,str,sum,super,tuple,type,unichr,unicode,vars,xrange,zip,apply,buffer,coerce,intern},%
    sensitive=true,%
    morecomment=[l]\#,%
    morestring=[b]',%
    morestring=[b]",%
    morestring=[s]{'''}{'''},
    morestring=[s]{"""}{"""},
    morestring=[s]{r'}{'},
    morestring=[s]{r"}{"},%
    morestring=[s]{r'''}{'''},%
    morestring=[s]{r"""}{"""},%
    morestring=[s]{u'}{'},
    morestring=[s]{u"}{"},%
    morestring=[s]{u'''}{'''},%
    morestring=[s]{u"""}{"""},%
    literate=
    {á}{{\'a}}1 {é}{{\'e}}1 {í}{{\'i}}1 {ó}{{\'o}}1 {ú}{{\'u}}1
    {Á}{{\'A}}1 {É}{{\'E}}1 {Í}{{\'I}}1 {Ó}{{\'O}}1 {Ú}{{\'U}}1
    {à}{{\`a}}1 {è}{{\`e}}1 {ì}{{\`i}}1 {ò}{{\`o}}1 {ù}{{\`u}}1
    {À}{{\`A}}1 {È}{{\'E}}1 {Ì}{{\`I}}1 {Ò}{{\`O}}1 {Ù}{{\`U}}1
    {ä}{{\"a}}1 {ë}{{\"e}}1 {ï}{{\"i}}1 {ö}{{\"o}}1 {ü}{{\"u}}1
    {Ä}{{\"A}}1 {Ë}{{\"E}}1 {Ï}{{\"I}}1 {Ö}{{\"O}}1 {Ü}{{\"U}}1
    {â}{{\^a}}1 {ê}{{\^e}}1 {î}{{\^i}}1 {ô}{{\^o}}1 {û}{{\^u}}1
    {Â}{{\^A}}1 {Ê}{{\^E}}1 {Î}{{\^I}}1 {Ô}{{\^O}}1 {Û}{{\^U}}1
    {œ}{{\oe}}1 {Œ}{{\OE}}1 {æ}{{\ae}}1 {Æ}{{\AE}}1 {ß}{{\ss}}1
    {ç}{{\c c}}1 {Ç}{{\c C}}1 {ø}{{\o}}1 {å}{{\r a}}1 {Å}{{\r A}}1
    {€}{{\EUR}}1 {£}{{\pounds}}1
    {^}{{{\color{ipython_purple}\^{}}}}1
    {=}{{{\color{ipython_purple}=}}}1
    {+}{{{\color{ipython_purple}+}}}1
    {*}{{{\color{ipython_purple}$^\ast$}}}1
    {/}{{{\color{ipython_purple}/}}}1
    {+=}{{{+=}}}1
    {-=}{{{-=}}}1
    {*=}{{{$^\ast$=}}}1
    {/=}{{{/=}}}1,
    literate=
    *{-}{{{\color{ipython_purple}-}}}1
     {?}{{{\color{ipython_purple}?}}}1,
    identifierstyle=\color{black}\ttfamily,
    commentstyle=\color{ipython_cyan}\ttfamily,
    stringstyle=\color{ipython_red}\ttfamily,
    keepspaces=true,
    showspaces=false,
    showstringspaces=false,
    rulecolor=\color{ipython_frame},
    frame=single,
    frameround={t}{t}{t}{t},
    framexleftmargin=6mm,
    numbers=none,
    numberstyle=\tiny\color{halfgray},
    backgroundcolor=\color{ipython_bg},
    basicstyle=\footnotesize,
    keywordstyle=\color{ipython_green}\ttfamily,
    aboveskip=1.2em,
    belowskip=1.2em,
}

\usepackage{listings}
\lstdefinelanguage{Fortran90}{
    morekeywords={Real,Complex,Intent},%
    emph={End,Subroutine,dp,in,Function,Implicit,None},%
    emphstyle={\color{mathematica_purple}},    
    sensitive=true,%
    morecomment=[l]\#,%
    morestring=[b]',%
    morestring=[b]",%
    morestring=[s]{'''}{'''},
    morestring=[s]{"""}{"""},
    morestring=[s]{r'}{'},
    morestring=[s]{r"}{"},%
    morestring=[s]{r'''}{'''},%
    morestring=[s]{r"""}{"""},%
    morestring=[s]{u'}{'},
    morestring=[s]{u"}{"},%
    morestring=[s]{u'''}{'''},%
    morestring=[s]{u"""}{"""},%
    identifierstyle=\color{black}\ttfamily,
    commentstyle=\color{slha_cyan}\ttfamily,
    stringstyle=\color{slha_red}\ttfamily,
    keepspaces=true,
    showspaces=false,
    showstringspaces=false,
    rulecolor=\color{slha_frame},
    frame=true,
    frameround={t}{t}{t}{t},
    framexleftmargin=6mm,
    numbers=none,
    numberstyle=\tiny\color{halfgray},
    backgroundcolor=\color{slha_bg},
    basicstyle=\footnotesize,
    keywordstyle=\color{slha_green}\ttfamily,
    aboveskip=1.2em,
    belowskip=1.2em,
}

\lstdefinelanguage{SLHA}{
    morekeywords={block,Block,BLOCK,decay,Decay,DECAY},%
    sensitive=true,%
    morecomment=[l]\#,%
    morestring=[b]',%
    morestring=[b]",%
    morestring=[s]{'''}{'''},
    morestring=[s]{"""}{"""},
    morestring=[s]{r'}{'},
    morestring=[s]{r"}{"},%
    morestring=[s]{r'''}{'''},%
    morestring=[s]{r"""}{"""},%
    morestring=[s]{u'}{'},
    morestring=[s]{u"}{"},%
    morestring=[s]{u'''}{'''},%
    morestring=[s]{u"""}{"""},%
    identifierstyle=\color{black}\ttfamily,
    commentstyle=\color{slha_cyan}\ttfamily,
    stringstyle=\color{slha_red}\ttfamily,
    keepspaces=true,
    showspaces=false,
    showstringspaces=false,
    rulecolor=\color{slha_frame},
    frame=true,
    frameround={t}{t}{t}{t},
    framexleftmargin=6mm,
    numbers=none,
    numberstyle=\tiny\color{halfgray},
    backgroundcolor=\color{slha_bg},
    basicstyle=\footnotesize,
    keywordstyle=\color{slha_green}\ttfamily,
    aboveskip=1.2em,
    belowskip=1.2em,
}

\lstdefinestyle{terminal} {
  morekeywords={cp,-r,make,cd},
  numbers=none, 
  stepnumber=1, 
  numberstyle=\tiny\color{halfgray}, 
  numbersep=10pt, 
  backgroundcolor=\color{black}, 
  basicstyle=\color{white}\ttfamily,
  stringstyle=\color{white}\ttfamily,
  keywordstyle=\color{white}\ttfamily\bfseries
 }

\lstdefinelanguage{MathIn}{
    morekeywords={Simplify,Eigenvalues,epsUV,Delta,UVscaleQ},%
    emph={Start,InitUnitarity,GetScatteringDiagrams,BuildScatteringMatrix,MakeSPheno,InitMatching,EFTcoupLO,EFTcoupNLO},%
    emphstyle={\color{mathematica_purple}},
    sensitive=true,%
    morecomment=[l]\%,%
    morestring=[b]',%
    morestring=[b]",%
    morestring=[s]{'''}{'''},
    morestring=[s]{"""}{"""},
    morestring=[s]{r'}{'},
    morestring=[s]{r"}{"},%
    morestring=[s]{r'''}{'''},%
    morestring=[s]{r"""}{"""},%
    morestring=[s]{u'}{'},
    morestring=[s]{u"}{"},%
    morestring=[s]{u'''}{'''},%
    morestring=[s]{u"""}{"""},%
    identifierstyle=\color{black}\ttfamily,
    commentstyle=\color{mathematica_cyan}\ttfamily,
    stringstyle=\color{mathematica_red}\ttfamily,
    keepspaces=true,
    showspaces=false,
    showstringspaces=false,
    rulecolor=\color{mathematica_frame},
    frame=none,
    numbers=left,
    numberstyle=\tiny\color{halfgray},
    basicstyle=\footnotesize,
    keywordstyle=\color{mathematica_green}\ttfamily,
    aboveskip=0.2em,
    belowskip=0.2em
}

\lstdefinelanguage{MathOut}{
    morekeywords={Simplify,Eigenvalues},%
    sensitive=true,%
    morecomment=[l]\%,%
    morestring=[b]',%
    morestring=[b]",%
    morestring=[s]{'''}{'''},
    morestring=[s]{"""}{"""},
    morestring=[s]{r'}{'},
    morestring=[s]{r"}{"},%
    morestring=[s]{r'''}{'''},%
    morestring=[s]{r"""}{"""},%
    morestring=[s]{u'}{'},
    morestring=[s]{u"}{"},%
    morestring=[s]{u'''}{'''},%
    morestring=[s]{u"""}{"""},%
    identifierstyle=\color{black}\ttfamily,
    commentstyle=\color{mathematica_cyan}\ttfamily,
    stringstyle=\color{mathematica_red}\ttfamily,
    keepspaces=true,
    showspaces=false,
    showstringspaces=false,
    rulecolor=\color{mathematica_frame},
    frame=none,
    frameround={t}{t}{t}{t},
    framexleftmargin=10mm,
    numbers=left,
    numberstyle=\tiny\color{halfgray},
    basicstyle=\footnotesize,
    keywordstyle=\color{mathematica_green}\ttfamily,
    aboveskip=0.2em,
    belowskip=0.2em,
}

\let\origthelstnumber\thelstnumber
\makeatletter
\newcommand*\Suppressnumber{%
  \lst@AddToHook{OnNewLine}{%
    \let\thelstnumber\relax%
     \advance\c@lstnumber-\@ne\relax%
    }%
}

\newcommand*\Reactivatenumber{%
  \lst@AddToHook{OnNewLine}{%
   \let\thelstnumber\origthelstnumber%
   \advance\c@lstnumber\@ne\relax}%
}

\def\postbreak{%
  \raisebox{0ex}[0ex][0ex]{\ensuremath{\hookrightarrow\space}}}

\def\thv[#1,#2,#3]{\left( \begin{array}{c} #1 \\ #2 \\ #3 \end{array} \right)}
\def\twv[#1,#2]{\left( \begin{array}{c} #1 \\ #2 \end{array} \right)}

\def\beq{\begin{equation}}
\def\eeq{\end{equation}}

\newcommand\SARAH{{\tt SARAH}\xspace}

\newcommand\SPheno{{\tt SPheno}\xspace}
\newcommand\Vevacious{{\tt Vevacious}\xspace}
\newcommand\MO{{\tt MicrOmegas}\xspace}
\newcommand\HB{{\tt HiggsBounds}\xspace}
\newcommand\HS{{\tt HiggsSignals}\xspace}

\newcommand\python{{\tt Python}\xspace}
\newcommand\pytorch{{\tt pytorch}\xspace}
\newcommand\xBIT{{\tt xBIT}\xspace}
\newcommand\github{{\tt GitHub}\xspace}

\definecolor{fsblue}{rgb}{0.,.0,1.}

\begin{document}

\title{\xBIT: an easy to use scanning tool with machine learning abilities}
\author{
   Florian Staub \thanksref{a1, a2} 
   }

\institute{
Institute for Theoretical Physics (ITP), Karlsruhe Institute of Technology, Engesserstra{\ss}e 7, D-76128 Karlsruhe, Germany \label{a1}
\and
Institute for Nuclear Physics (IKP), Karlsruhe Institute of Technology, Hermann-von-Helmholtz-Platz 1, D-76344 Eggenstein-Leopoldshafen, Germany \label{a2}
}


\maketitle

\begin{abstract}
\xBIT is a tool for performing parameter scans in beyond the Standard Model theories. It's written in \python and fully open source. The main purpose of \xBIT is to provide an easy to use tool to help phenomenologists with their daily task: exploring the parameter space of new models. It was developed under the impression of the \SARAH/\SPheno framework, but should be use-able with other tools as well
that use the SLHA format to transfer data. It also supports by default \MO for dark matter calculations, \HB and \HS for checking the Higgs properties, and \Vevacious for testing the vacuum stability. Classes for other tools can be added if necessary. In order to improve the efficiency of the parameter scans, 
the recently proposed 'Machine Learning  Scan' approach is included. For this purpose, \xBIT uses \pytorch to deal with artificial neural networks. 
\end{abstract}

\section{Introduction}
\label{sec:intro}
The second run of the Large Hadron Collider (LHC) has finished, but the standard model (SM) of particle physics is still the best model we have to explain all the data. The pile up of null results in the searches for new physics has 
significantly changed the attitude of the community. While there was one clear favourite for the successor of the SM, namely the minimal supersymmetric 
standard model (MSSM), before the LHC has been started, the situation is very unclear nowadays. Consequently, many more models for physics beyond the standard model (BSM) are intensively studied today. All of these models have in common that 
they must be confronted with experimental and theoretical constraints at a given point. That means, one needs to investigate the parameter space of these models using numerical tools. Very often one is mainly interested in finding parameter regions which are in agreement with all constraints. For this purpose, one doesn't need all the overload necessary for statistical interpretations of such regions. \\
In this paper a new tool called \xBIT is introduced which was created for exploring the parameter space in a given theory: it shall provide an easy and fast possibility to perform scans in a new model using well established tools like \SARAH\cite{Staub:2008uz,Staub:2009bi,Staub:2010jh,Staub:2012pb,Staub:2013tta}/\SPheno\cite{Porod:2003um,Porod:2011nf}, \MO, \HB\cite{Bechtle:2008jh,Bechtle:2011sb,Bechtle:2013wla}, \HS \cite{Bechtle:2013xfa} and \Vevacious \cite{Camargo-Molina:2013qva}. The motivation to use \SARAH/\SPheno as basis for \xBIT is obvious: this is the only combination of tools which provides all information which is necessary to study a new model, e.g. scalar masses at the two-loop level \cite{Goodsell:2015yca,Goodsell:2015ira,Braathen:2017izn}, decay widths at tree- and one-loop-level \cite{Goodsell:2017pdq}, predictions for the most important flavour and precision observables \cite{Porod:2014xia}, checks of the tree-level unitarity constraints including the $s$-dependence \cite{Goodsell:2018tti}. \\
\xBIT supports simple grid and random scan, but also more sophisticated approaches to efficiently scan parameter spaces are available. In particular, the machine learning scan (MLS) algorithm \cite{Ren:2017ymm} based on artificial neural networks has been implemented. Moreover, also new classes for other tools or scans can be added by the user if necessary. \\
This paper is organised as follows: \cref{sec:installation} explains the installation of \xBIT. How a scan can be defined and performed is explained in \cref{sec:setup} before it is shown how to run \xBIT in \cref{sec:running}. An example how the different scan options can be used is given \cref{sec:example}. A summary with some additional remarks is given in \cref{sec:summary}. \cref{sec:scans} gives a short introduction to the supported types of scans, while \cref{app:classes} explains the implementation of new classes for tools and scans.
\section{Installation}
\label{sec:installation}
\xBIT as available at \github:
\begin{center}
{\tt https://github.com/fstaub/xBIT}
\end{center}
One can simply download the {\tt zip} file and extract it or clone the repository. For more details how to handle the repositories, please, check the \github documentation. In order to run \xBIT, it's recommend to use \python 3\footnote{Some efforts were made to make \xBIT 
compatible with \python 2 but this was not tested in detail.}. Moreover, the following (standard) packages must be installed: {\tt numpy}, {\tt six}, {\tt curses}, {\tt os}, {\tt time}, {\tt sys}, {\tt shutil}, {\tt collections}. Moreover, {\tt xSLHA} \cite{Staub:2018rih} is used to read SLHA files. {\tt xSLHA} can be installed globally via 
\begin{lstlisting}[style=terminal]
> sudo pip3 install xslha
\end{lstlisting}
or, without root access, via 
\begin{lstlisting}[style=terminal]
> pip3 install --user xslha
\end{lstlisting}
Finally, \pytorch is needed for using artificial neural networks. Detailed instructions for installing it are given on the homepage of \pytorch:
\begin{center}
 {\tt https://pytorch.org/}
\end{center}

\section{Setup}
\label{sec:setup}
In order to run a scan, two input files are needed:
\begin{itemize}
 \item {\bf The settings file}: this file contains the information about the (local) installation of the tools which should be used during the scan. This file can be used for all scans in the same model.
 \item {\bf The scan file}: this file defines an actual scan: included tools, parameter ranges, used options for the different scan types, demanded values of observables, and so on .
\end{itemize}

\subsection{Settings File}
\label{sec:settings}
This file contains all necessary information to run a code. 
An example for the MSSM might look like:
\begin{lstlisting}[language=ipython]
{
  "SPheno": {
    "Command": "~/HEP_Tools/SPheno-4.0.3/bin/SPhenoMSSM",
    "InputFile": "LesHouches.in.MSSM",
    "OutputFile": "SPheno.spc.MSSM"
  },

  "MicrOmegas": {
     "Command": "~/HEP_Tools/micromegas_5.0.2/MSSM_SARAH/CalcOmega",
     "OutputFile":  "omg.out",
     "DM_Candidate":  1000022
  },

  "HiggsBounds": {
    "Command": "~/HEP_Tools//HiggsBounds-4.3.1/HiggsBounds",
    "Options": "LandH effC 3 1",
    "OutputFile": "HiggsBounds_results.dat"
  },

  "HiggsSignals": {
    "Command": "~/HEP_Tools/HiggsSignals-1.3.2/HiggsSignals",
    "Options":"latestresults peak 2 effC 3 1",
    "OutputFile": "HiggsSignals_results.dat"
  },

  "Vevacious": {
    "Path": "~/HEP_Tools/Vevacious-1.1.2/",
    "Command": "./bin/Vevacious.exe --input=bin/VevaciousInitialization_stop.xml --should_Tunnel=False"
  }  

} 
\end{lstlisting}
The properties of all the tools but \SPheno are defined via the source code files located in {\tt package/tools/}.  
If another tool shall be used, it's sufficient to add a new \python file which defines the class {\tt NewTool} for this tool. 
Moreover, the string defined via {\tt self.name} in this class is the one which 
is also used in the settings file to identify the tool. See \cref{app:tools} for more details have to define a new class to include other tools. 

\subsection{Scan File}
\label{sec:scanfile}
All properties of the scan are defined in another {\tt json} file which is usually located in the sub-directory {\tt Input}. 
This file contains the information described in the following subsections. 

\subsubsection{General Options}
The block {\tt Setup} is used to set define the basic properties of a scan:
\begin{lstlisting}[language=ipython]
  "Setup": {
    "Settings": STRING,
    "Name": STRING,
    "Type": STRING,
    "Points": INT,
    "Iterations": INT,
    "Cores": INT
  },
\end{lstlisting}
The meaning of the different entries is:
\begin{itemize}
 \item {\tt Settings}: name of the file containing the settings for the different tools 
 \item {\tt Name}: a name for the scan
 \item {\tt Type}: The type of scan. Up to now the following types are defined: 
  \begin{itemize}
   \item {\tt Grid}
   \item {\tt Random}
   \item {\tt MCMC}
   \item {\tt MLS}
  \end{itemize}
  See \cref{sec:scans} for more details.
 \item {\tt Points}: the number of scan points. Note:
  \begin{itemize}
   \item This entry has no effect for grid scans 
   \item For a MLS scan this defines the number of scan points {\it per iteration}. In The first iteration a bigger sample with five times that number of points is generated. 
  \end{itemize}
 \item {\tt Iterations}: number of iterations (only for MLS scans)
 \item {\tt Cores}: number of used cores (the package {\tt multiprocess} is used for parallelisation)
\end{itemize}

\subsubsection{Included Tools}
One can decide which tools shall be included in the scan. All tools which are defined in {\tt package/tools/}, as mentioned in \cref{app:tools}, can be turned on or off. Currently, this means
for the public version of \xBIT:
\begin{lstlisting}[language=ipython]
  "Included_Codes": {
    "HiggsBounds": "True" or "False",
    "HiggsSignals": "True" or "False",
    "MicrOmegas": "True" or "False",
    "Vevacious": "True" or "False"
  } 
\end{lstlisting}

\subsubsection{Observables}
One can define the position of observables in the SLHA files together with their best fit value and variance. This information is used by \xBIT to calculate the likelihood of a point by assuming an Gaussian distribution. 

\begin{lstlisting}[language=ipython]
  "Observables": {
      "0": {
        "SLHA": [BLOCK, ENTRY],
        "MEAN": FLOAT,
        "VARIANCE": FLOAT
      },
     "1": {
       ...
       },
     ...  
  } 
\end{lstlisting}
{\it Example:} The Higgs mass and dark matter relic density are set by
\begin{lstlisting}[language=ipython]
  "Observables": {
      "0": {
        "SLHA": [MASS, [25]],
        "MEAN": 125.,
        "VARIANCE": 1.
      },
     "1": {
        "SLHA": [DARKMATTER, [1]],
        "MEAN": 0.1,
        "VARIANCE": 0.01.
       }
  } 
\end{lstlisting}

\subsubsection{Machine Learning / Neural Network}
For the MLS approach one or two deep neural network are used. The properties of the network as well as of the training period is defined in the block {\tt ML}. 
\begin{lstlisting}[language=ipython]
  "ML": {
    "Neurons": ARRAY,
    "LR": FLOAT,
    "Classifier": "True" or "False",
    "DensityPenality": "False"  or "False",
    "TrainLH": "True"  or "False",
    "Epochs": INT
  }
\end{lstlisting}
The purpose of the different entries is:
\begin{itemize}
 \item {\tt Neurons}: the number of neurons per hidden layer. For instance, to use three hidden layers with 25 neuron in each layer
\begin{lstlisting}[language=ipython]
"Neurons": [25,25,25]
\end{lstlisting}
 There is no constraint on the number of hidden layers 
 \item {\tt LR}: the learning rate used by the Adam optimiser, e.g.
\begin{lstlisting}[language=ipython]
"LR": 0.001
\end{lstlisting}
 \item {\tt "Epochs"}: the maximal number of epochs used in the training of the neural network
 \item {\tt "Classifier"}: defines if a classifier to filter valid/invalid points should be included. Invalid points are defined as:
 \begin{itemize}
  \item No spectrum file has been produced 
  \item Not all observables could be calculated (for instance because of a charged LSP)
 \end{itemize}
 \item {\tt TrainLH}: defines if the neural network should be trained with the likelihood or with the individual observables 
 \item {\tt DensityPenality}: defines if a penalty for points too close to already scanned points should be included to prevent an oversampling of parameter regions. See \cref{sec:MLS}
\end{itemize}

\subsubsection{Scan Parameters}
The parameters which shall be varied are defined in  {\tt Variables} together with the chosen distributions. 
\begin{lstlisting}[language=ipython]
   "Variables": {
      "0": "FUNCTION",
      "1": "FUNCTION",
      ...
  }
\end{lstlisting}
Here, {\tt FUNCTION} is usually a string calling a {\tt numpy} ({\tt np}) routine, e.g. for random variables
\begin{lstlisting}[language=ipython]
   "Variables": {
      "0": "np.random.uniform(100,1000)",
      "1": "np.exp(np.random.uniform(np.log(0.01),np.log(100)))"
  }
\end{lstlisting}
and 
\begin{lstlisting}[language=ipython]
   "Variables": {
      "0": "np.linspace(100, 1000, num=50)",
      "1": "np.logspace(np.log(0.01),np.log(100), num=50)"
  }
\end{lstlisting}
for a grid scan. See also the discussion in \cref{sec:scans}. \\
These variables can be used then in the block containing the input parameters for the different tools as explained next.

\subsubsection{Input Blocks}
Finally, we can define what is written to the Les Houches input files for the spectrum generator. This is defined in the block {\tt Blocks}:
\begin{lstlisting}[language=ipython]
"Blocks": {
    "BLOCKNAME1": {
         "1": VALUE,
         "2": VALUE,
         "3": VALUE,
         ...
    },

    "BLOCKNAME2": {
         "1 1": VALUE,
         "1 2": VALUE,
         "1 3": VALUE,
         "2 1": VALUE,
         ...
    },
    ...
  }  
\end{lstlisting}
Thus, the definition is completely agnostic about SLHA conventions and all input blocks necessary for the considered model can be listed. Possible values for each entry are
\begin{itemize}
 \item {\bf integer}
\begin{lstlisting}[language=ipython]
"1": 1
\end{lstlisting}
 \item {\bf float}
\begin{lstlisting}[language=ipython]
"1": 0.01
\end{lstlisting} 
 \item {\bf Scan variable}
\begin{lstlisting}[language=ipython]
"1": "Variable[1]"
\end{lstlisting}
  \item {\bf A function of scan variables}
\begin{lstlisting}[language=ipython]
"1": "2*Variable[1]*Variable[2]"
\end{lstlisting}
\end{itemize}
\paragraph*{Example:} If one would like -- why-ever --to define  a scan with the following properties: 
\begin{itemize}
 \item a log distribution for $m_0 = [10^2,10^5]$~\text{GeV} (100 steps)
 \item a linear distribution for $\tan\beta=[5,50]$ (46 steps)
 \item $M_{12} = 2 m_0$ 
 \item $A_0 = m_0/\tan\beta$
 \item $\mu >0 $
\end{itemize}
that's done via
\begin{lstlisting}[language=ipython]
{
...
"Variables": {
      "0": "np.geomspace(100, 100000, num=100)",
      "1": "np.linspace(5,50, num=46)"
    },
"Blocks": {
    "MINPAR": {
         "1": "VARIABLE[0]",
         "2": "2*VARIABLE[0]",
         "3": "VARIABLE[1]",
         "4": 1.0,
         "5": "VARIABLE[0]/VARIABLE[1]"
    },
  ...
  },
   ...
} 
\end{lstlisting}

\section{Running \xBIT}
\label{sec:running}
Once all files are at place, a scan is executed via
 \begin{lstlisting}[style=terminal]
 > python3 xBIT.py INPUTFILE --Options
 \end{lstlisting}
The following options are possible to steer the run and screen output:
\begin{itemize}
 \item[]{\tt --short}: store output in short form, i.e. keeping only valid points
 \item[]{\tt --debug}: print debug information on screen 
 \item[]{\tt --clean}: clean up the temporary directory before running 
 \item[]{\tt --quiet}: use minimal screen output 
 \item[]{\tt --curses}: use the {\tt curses} package to show more information about the progress on the screen
\end{itemize}
In addition, settings used in the given input file can be overwritten:
\begin{itemize}
 \item[]{\tt --Name=STRING}: sets the name of the scan 
 \item[]{\tt --Points=INT}: sets the name of points
 \item[]{\tt --Cores=INT}: sets the name of used cores
 \item[]{\tt --Iterations=INT}: number of iterations for a MLS scan
 \end{itemize}
Moreover, the properties of the neural network can be modified compared to the ones defined in the input file:
\begin{itemize}
 \item[]{\tt --Classifier=BOOL}: include/exclude classifier 
 \item[]{\tt --Epochs=INT}: number of epochs
 \item[]{\tt --LR=FLOAT}: learning rate of the neural network (Adam optimiser)
 \item[]{\tt --Neurons=STRING}: number of neurons in hidden layers, e.g. \verb"\"[25,25,25]\""
 \item[]{\tt --TrainLH=BOOL}: train directly the likelihood instead of the observables
 \item[]{\tt --DensityPenality=BOOL}: include a penalty for points to close to already sampled ones
 \end{itemize}
 The output is written to the sub-directory 
 \begin{center}
 {\tt Output/Name}
 \end{center}
 All spectrum files including the information of the other tools (\MO, \HB/\HS,\Vevacious) are combined to one big file. The individual parameter points are separated by the string \verb"ENDOFPARAMETERPOINT". \\
  For each scan also a temporary directory is created under 
 \begin{center} 
  {\tt Temp}
  \end{center}
  the name for the sub-directory of the current is the output of {\tt time.time()} when the scan was started.  This directory contains also log files with additional information that might be helpful for debugging.

\section{Example: Dark Matter in the constrained MSSM (CMSSM)}
\label{sec:example}
This section contains a well-known example to demonstrate how to use \xBIT to perform parameter scans: the dark matter relic density in the $m_0$/$M_{12}$ plane in the CMSSM. Since this is widely discussed topic in literature 
(see e.g. \cite{Ellis:2012aa}), all introduction concerning the underlying physics is skipped here. The focus is one the technical part: different parameter scans using 10,000 points in the following $2d$-plane shall be performed:
\begin{align}
 m_0 = & [100,2500]~\text{GeV} \\
 M_{1/2} = & [100,2500]~\text{GeV} \\
 A_0 = & 0\\
 \tan\beta = 10 \\
 \mu > 0 
\end{align}
In order to keep this example as simple as possible we completely ignore the Higgs mass constraint. Thus, the only experimental constraints which we include is the relic density 
$\Omega h^2 \sim 0.1$. It is well known that only fine-tuned regions in the CMSSM are consistent with this value. Therefore, it's interesting to check the results of the different scan types.

\subsubsection*{Grind and Random Scan}
\begin{figure}[tb]
\includegraphics[width=0.49\linewidth]{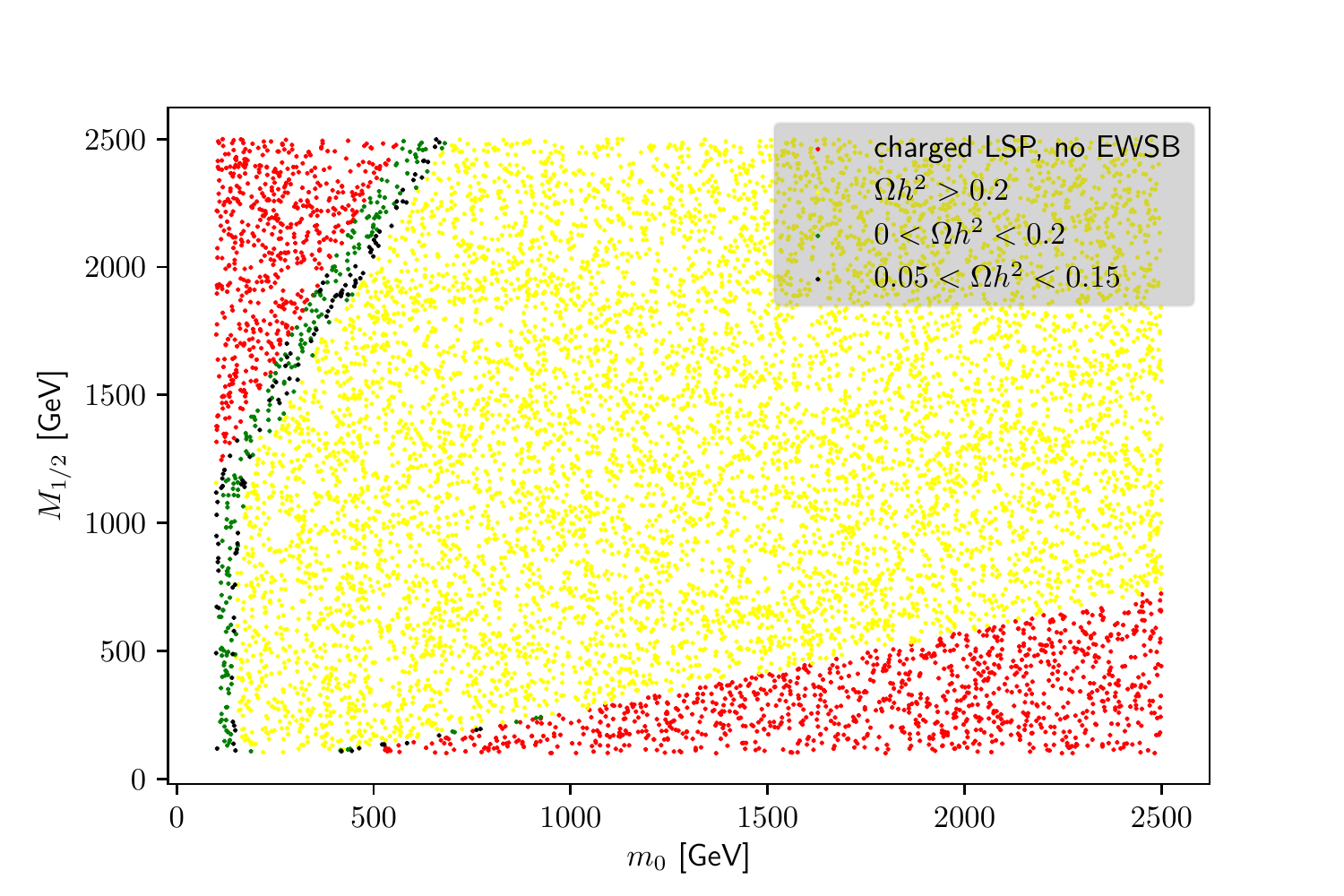} \hfill
\includegraphics[width=0.49\linewidth]{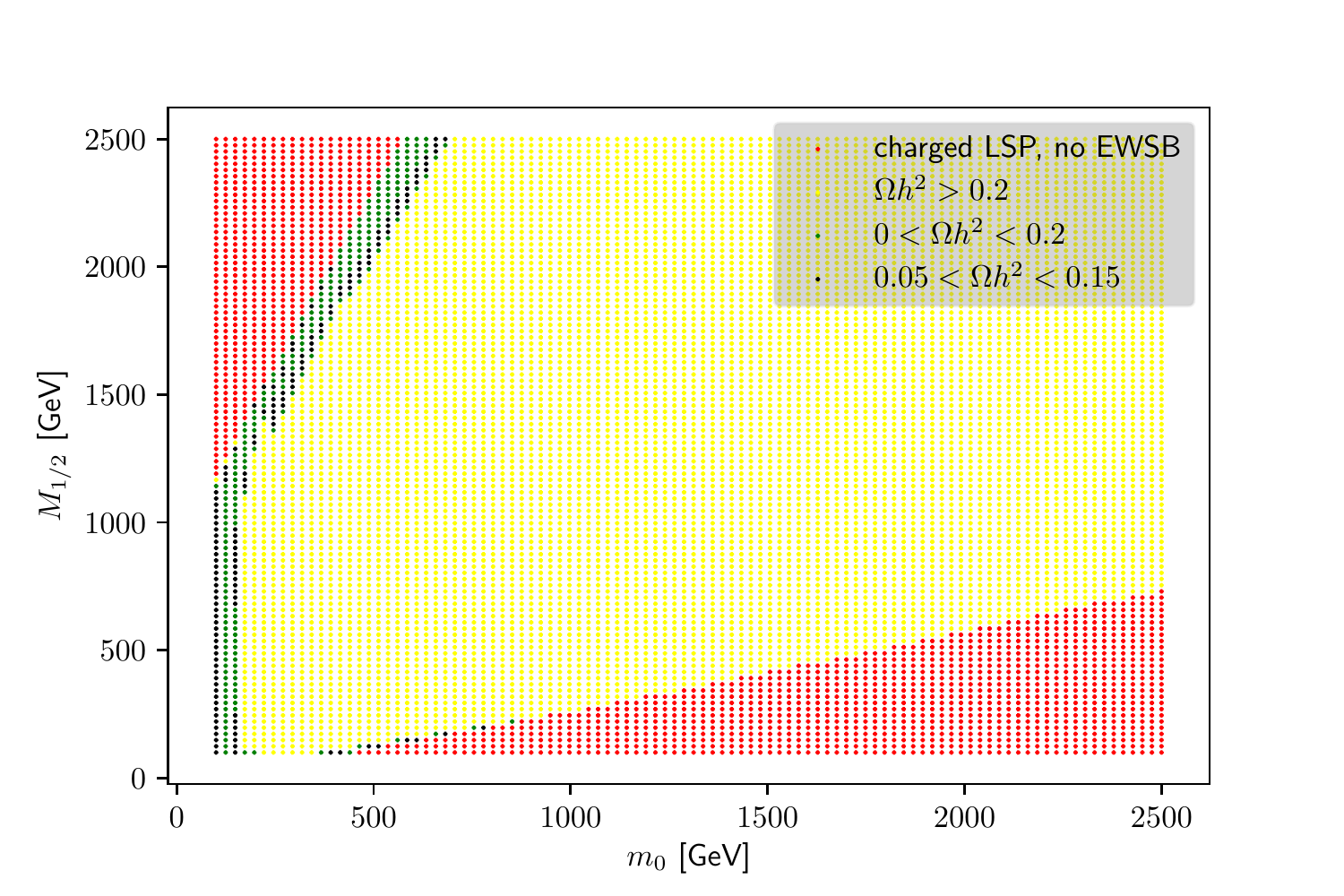}
\caption{The relic density in the $m_0/M_{12}$ plane in the CMSSM using either a random (on the left) or grid (on the right) scan. The red points are nonphysical (no EWSB or charged LSP) while yellow points have dark matter abundance which is much too high.}
\label{fig:grid_random}
\end{figure}

The two simplest options to scan the parameter space of a model is to use either a grid or random sampling. The relevant parts of the input file for \xBIT for the random scan are:
\begin{lstlisting}[language=ipython]
{
  "Setup": {
    "Settings": "MSSM.json",
    "Name": "m0_m12_DM_Random",
    "Type": "Random",
    "Points": 10000,
    "Cores": 1
  },

  "Included_Codes": {
    "MicrOmegas": "True",
    ...
  },

  "Variables": {
      "0": "np.random.uniform(100, 2500)",
      "1": "np.random.uniform(100, 2500)"
  },

...

   "MINPAR": {
         "1": "Variable[0]",
         "2": "Variable[1]",
         "3": 10.0,
         "4": 1.0,
         "5": 0.0
    },

\end{lstlisting}
while the grid scan is defined via 
\begin{lstlisting}[language=ipython]
{
  "Setup": {
    "Settings": "MSSM.json",
    "Name": "m0_m12_DM_Grid",
    "Type": "Grid",
    "Cores": 1
  },

  "Included_Codes": {
    "MicrOmegas": "True",
    ...
  },

  "Variables": {
      "0": "np.linspace(100, 2500, num=100)",
      "1": "np.linspace(100, 2500, num=100)"
  },

...
\end{lstlisting}
Note, in order to improve the speed of the scan one could have used more cores per scan if available. However, we are later interested in comparing the efficiencies of the different types of scans. Thus, we use always one core only. The results are shown in \cref{fig:grid_random}. One can see at this example that the efficiency of both scans is not very high, i.e. most points which were sampled have a relic density which is far away from the experimental value.

\subsubsection*{Machine learning scan}
\begin{figure}[tb]
\includegraphics[width=0.49\linewidth]{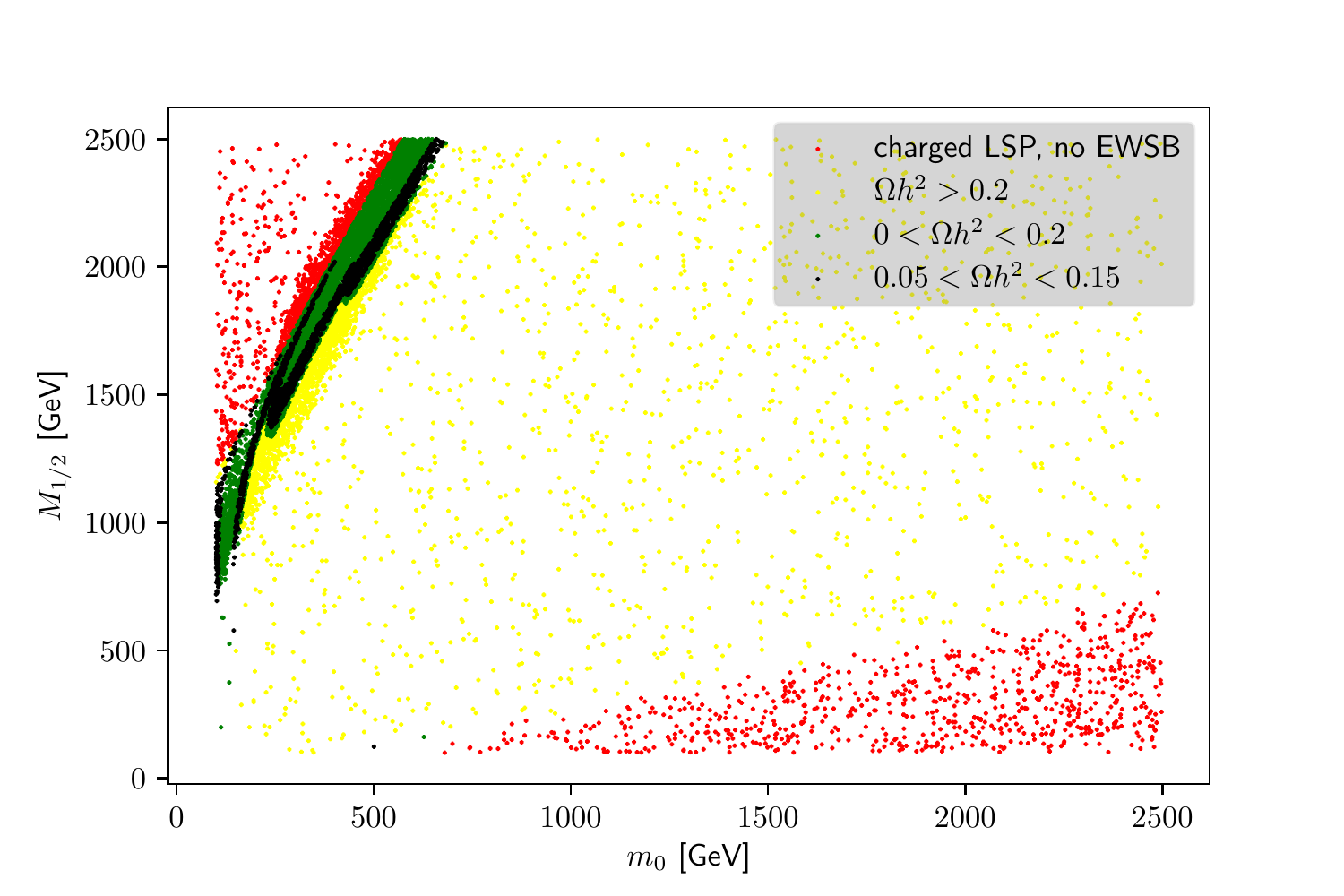} \hfill
\includegraphics[width=0.49\linewidth]{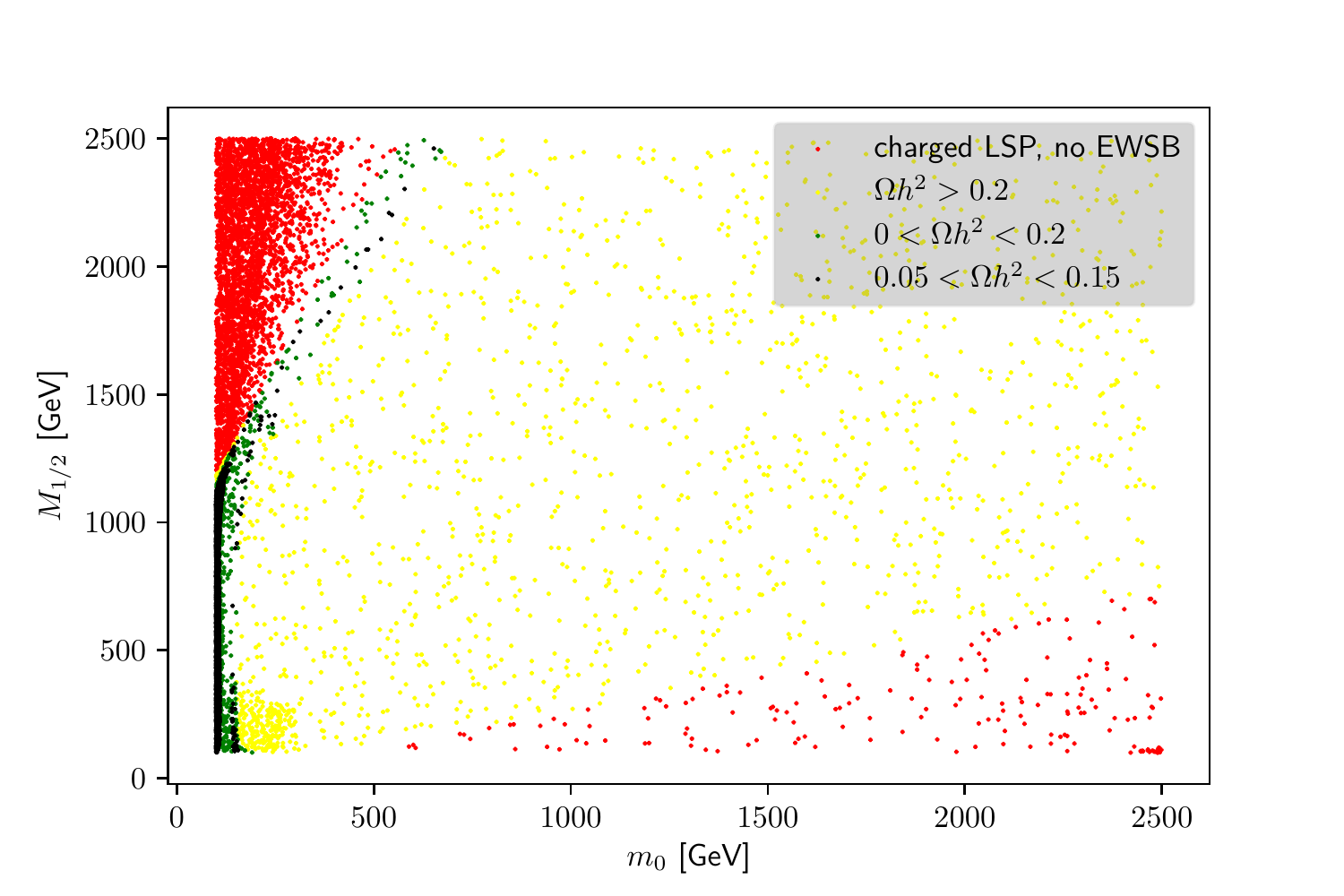}
\caption{The relic density in the $m_0/M_{12}$ plane in the CMSSM using the MLS approach with one neutral network as predictor. On the left the neural network is trained to predict the relic density while it is trained on the likelihood on the right. The colour code is the same as in \cref{fig:grid_random}.}
\label{fig:predictor}
\end{figure}

We turn now to the MLS approach which uses neutral networks. The simplest option is to use only one neural network as predictor. This neural network can either be trained to predict the value of the observables for each point or to predict 
the entire likelihood. We are going to check both options in the following. The entries in the input file which are different compared to the random scan are the following:
\begin{lstlisting}[language=ipython]
{
  "Setup": {
    "Settings": "MSSM.json",
    "Name": "m0_m12_DM_LH" or "m0_m12_DM_Obs",
    "Type": "MLS",
    "Points": 100,
    "Iterations": 100,
    "Cores": 1
  },

  "Observables": {
      "0": {
        "SLHA": ["DARKMATTER",[1]],
        "MEAN": 0.1,
        "VARIANCE": 0.01
      }
  },

  "ML": {
    "Neurons": [25,25,25],
    "LR": 0.001,
    "Classifier": "False",
    "DensityPenality": "False",
    "TrainLH": "True" or "False",
    "Epochs":5000
  }, 
 
\end{lstlisting}
We use here a neural network with three hidden layers with 25 neurons per layer. As already said, we use as training data either the overall likelihood of the point ({\tt TrainLH: True}) or the values of the individual observables 
({\tt TrainLH: False}). Since we have only one observable this results in the same topology of the neural network. However,  if several observables are defined, the training of the observables results in a bigger neural network and training might be significantly slower. The results are summarised in \cref{fig:predictor}. One can see that a problem with this scan is that it can happen that nonphysical regions are highly populated.  \\

\begin{figure}[tb]
\includegraphics[width=0.49\linewidth]{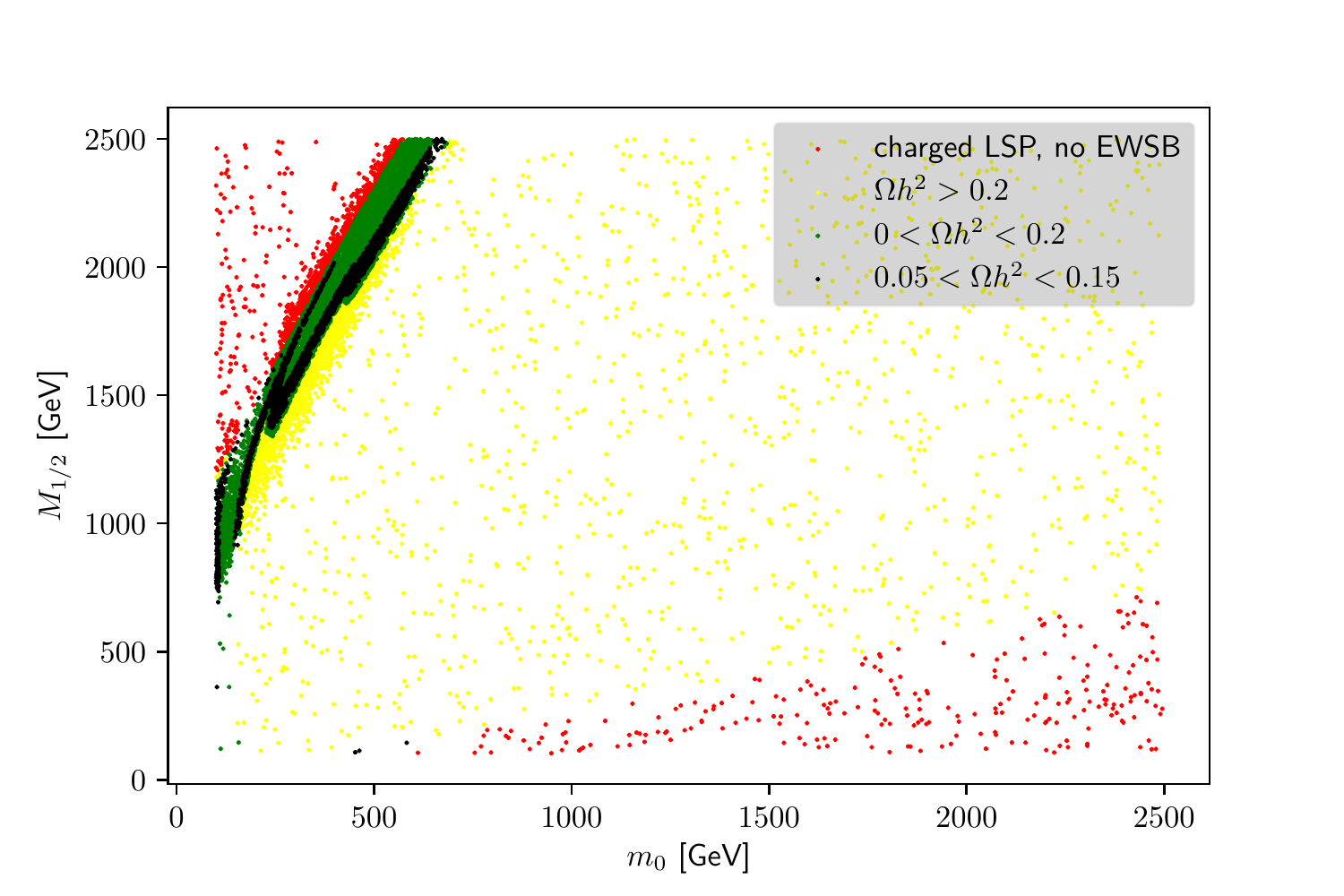} \hfill
\includegraphics[width=0.49\linewidth]{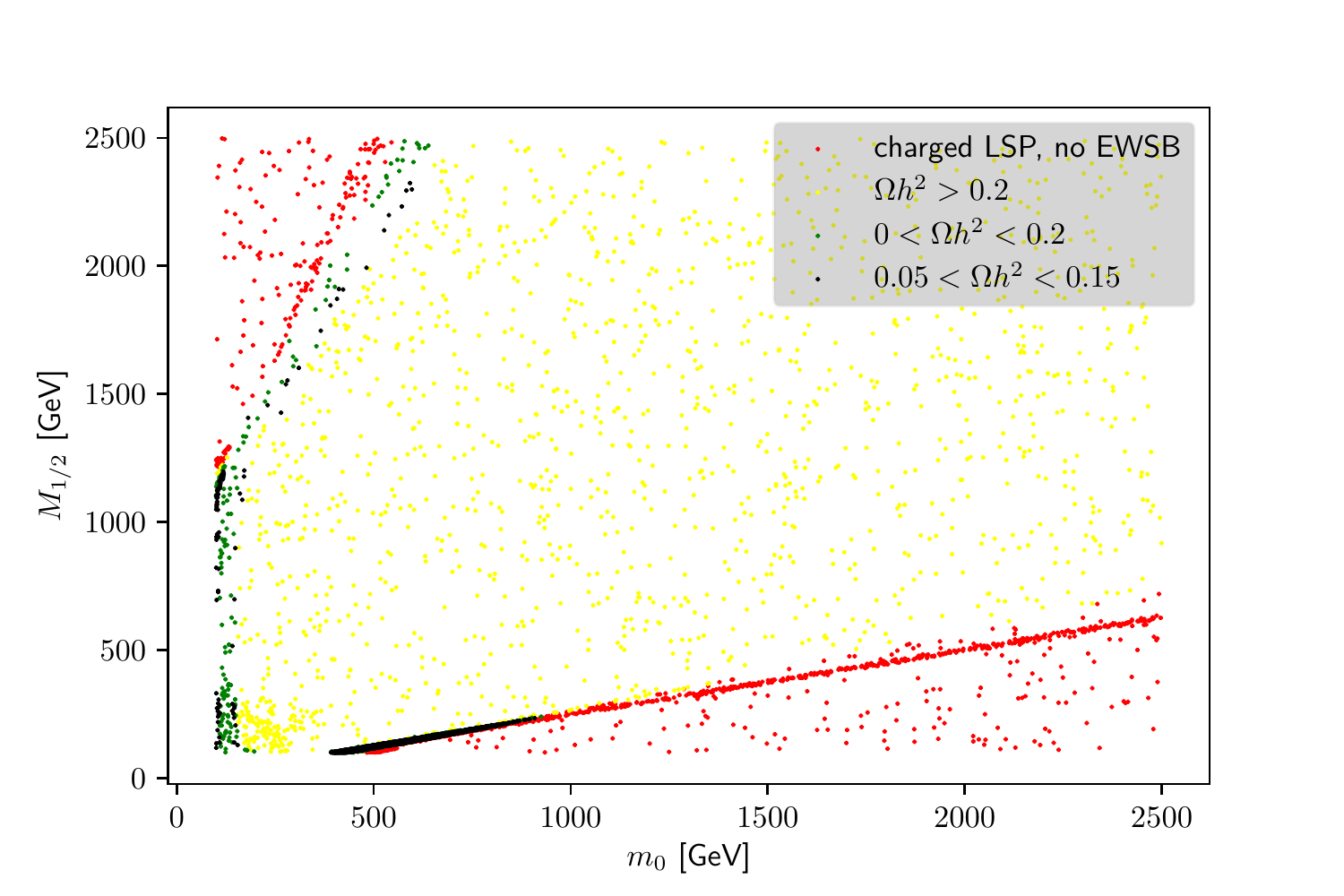}
\caption{The relic density in the $m_0/M_{12}$ plane in the CMSSM using the MLS approach with a predictor and classifier. On the left the neural network is trained to predict the relic density while it is trained on the likelihood on the right. The colour code is the same as in \cref{fig:grid_random}.}
\label{fig:classifier}
\end{figure}

In order to avoid the sampling of nonphysical points, one can turn on a second neutral network which classifies points into 'valid' and 'not valid' and drops all non valid points from the list of candidates for new scan points. This is done via:
\begin{lstlisting}[language=ipython]
  "ML": {
    "Neurons": [25,25,25],
    "LR": 0.001,
    "Classifier": "True",
    "DensityPenality": "False",
    "TrainLH": "True" or "False",
    "Epochs":5000
  }
\end{lstlisting}
The classifier network uses the same hyperparameter as the predictor. The impact of the second network is shown in \cref{fig:classifier}. One can see that this works very well and the number of nonphysical points is highly reduced. On the other side, it seems that the density of points in specific regions is very high while other promising regions are hardly sampled. One can also try to overcome this problem by either starting several scans in order to increase the probability to scan more parameter regions, or by introducing a penalty for regions with a large density. \\

\begin{figure}[tb]
\includegraphics[width=0.49\linewidth]{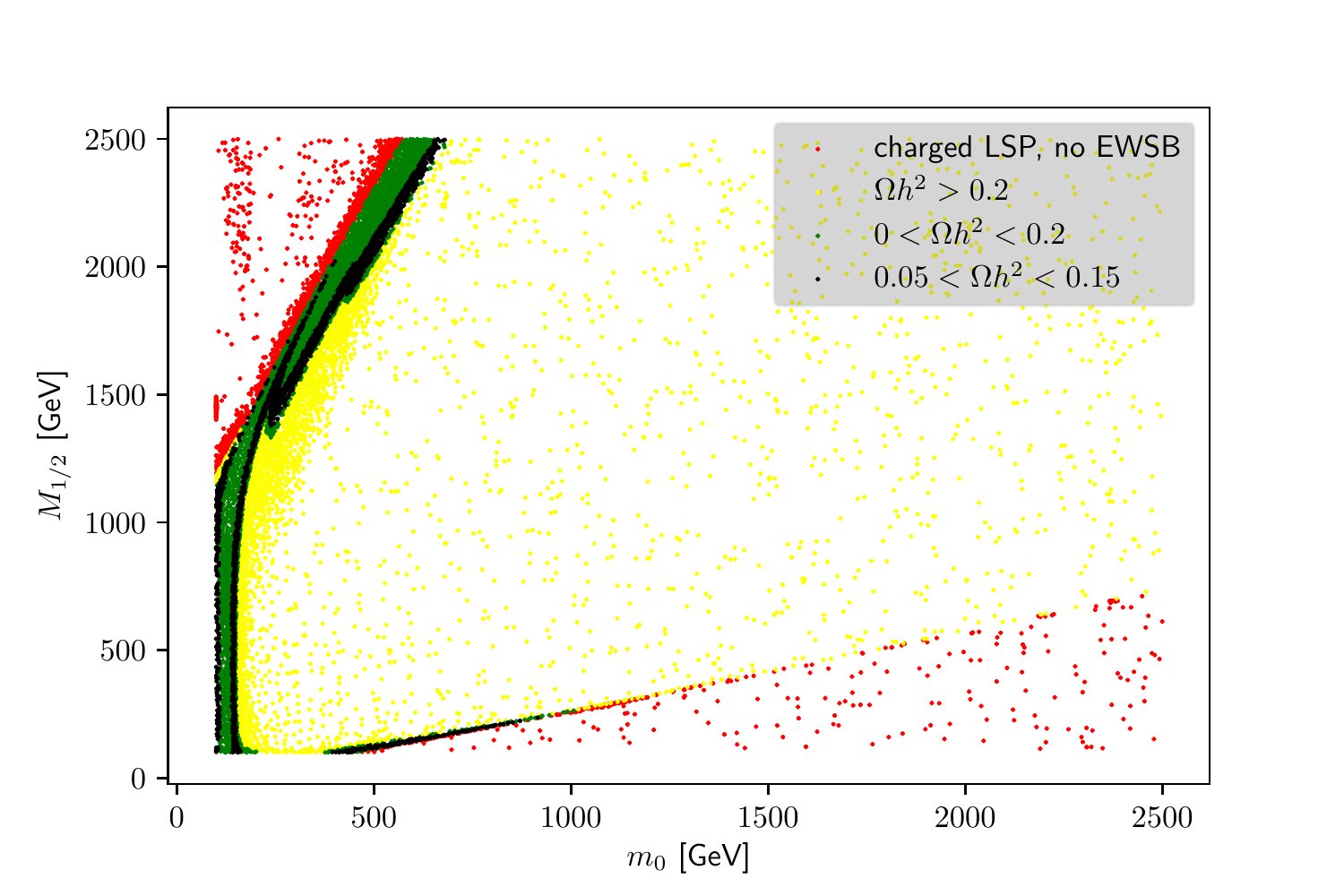} \hfill
\includegraphics[width=0.49\linewidth]{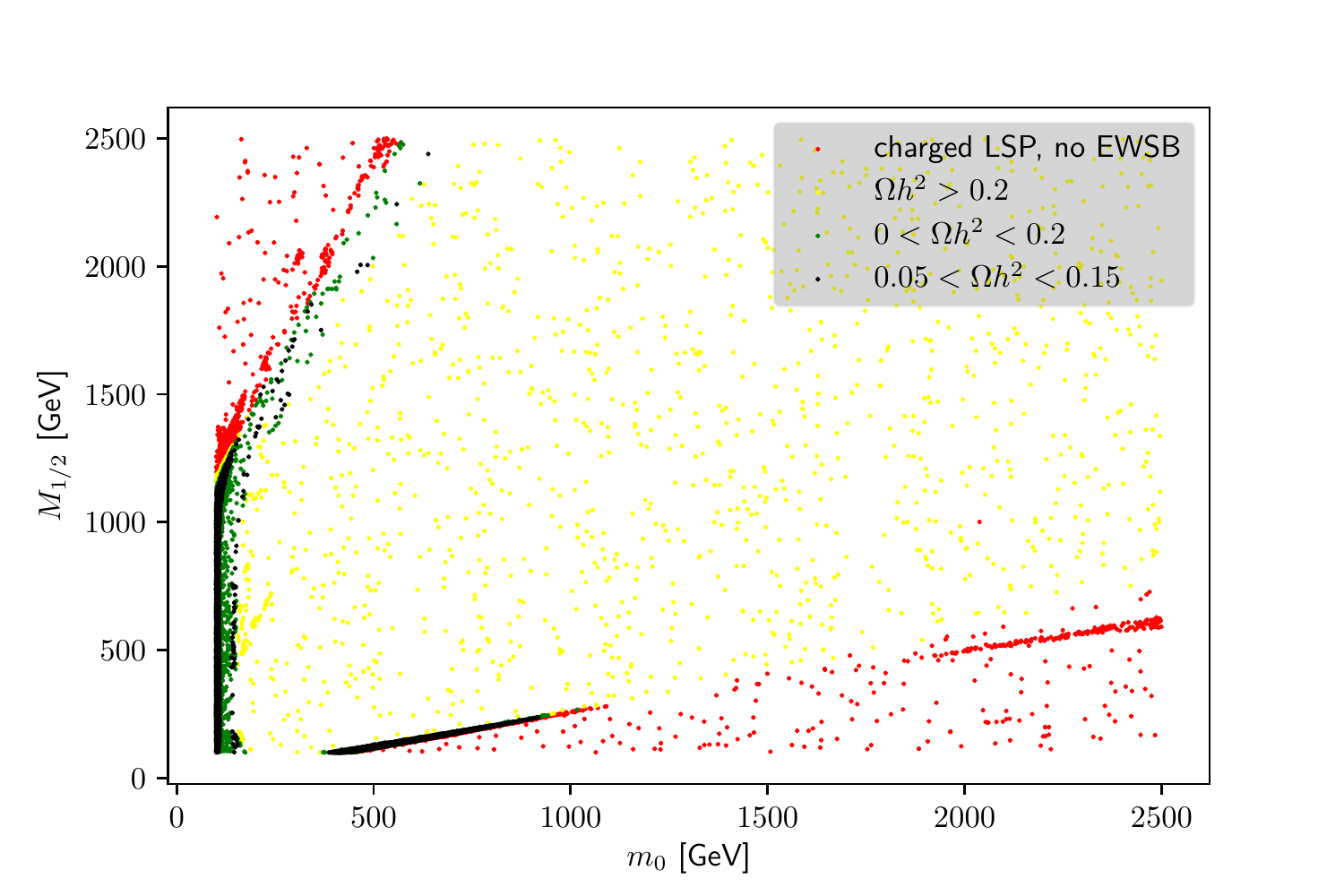}
\caption{The relic density in the $m_0/M_{12}$ plane in the CMSSM using the MLS approach with a predictor, a classifier and a density penalty. On the left the neural network is trained to predict the relic density while it is trained on the likelihood on the right. The colour code is the same as in \cref{fig:grid_random}.}
\label{fig:penality}
\end{figure}

As explained in \cref{sec:MLS} this penalty reduces the probability that a point is proposed which is very close to an already existing point. It's turned on via
\begin{lstlisting}[language=ipython]
  "ML": {
    "Neurons": [25,25,25],
    "LR": 0.001,
    "Classifier": "True",
    "DensityPenality": "True",
    "TrainLH": "True" or "False",
    "Epochs":5000
  }
\end{lstlisting}
The outcome is shown in \cref{fig:penality}. One can see that this works nicely especially for the case that the network is trained to predict the relic density. The impact in the case that the network is trained to predict the likelihood is much smaller. The reason is most likely that the likelihood varies much faster than the relic density. Thus, one might need also a penalty function with some exponential behaviour. However, this is left for further investigation.   \\

\begin{figure}[tb]
\includegraphics[width=0.49\linewidth]{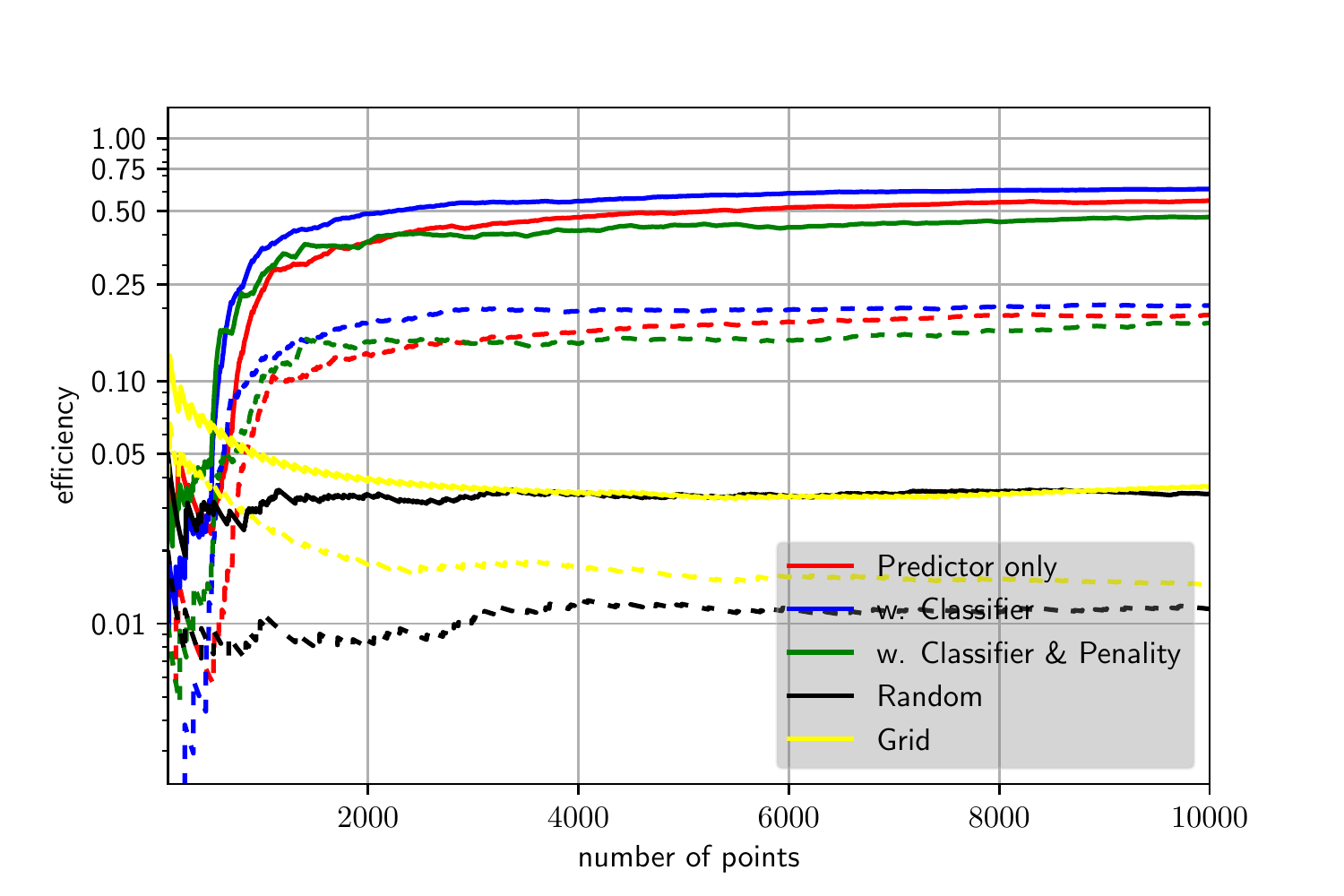} \hfill
\includegraphics[width=0.49\linewidth]{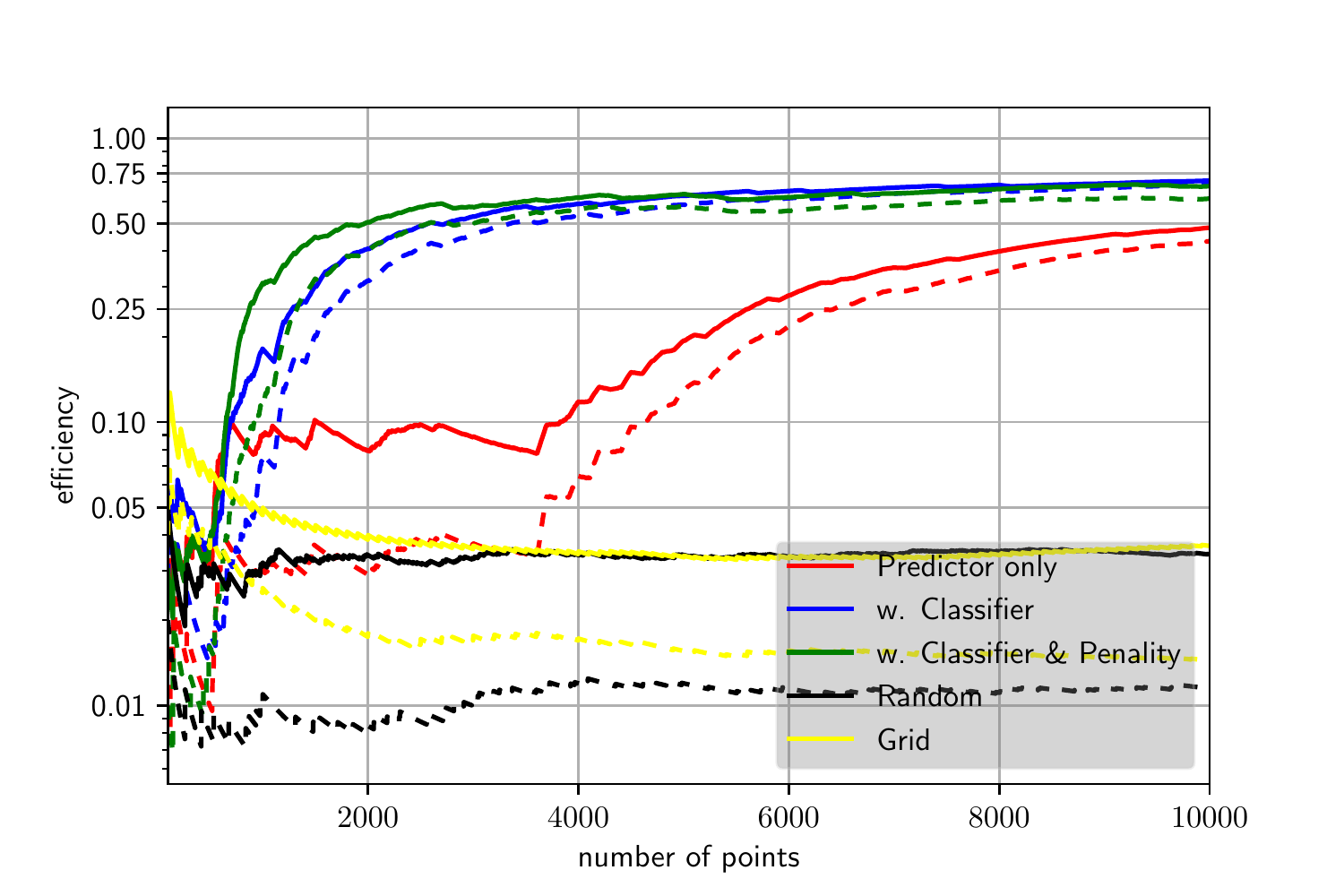} 
\caption{The efficiency of the dark matter scans of the different types of scans. On the left the neural network is trained to predict the relic density while it is trained to predict the entire likelihood on the right. The full lines correspond to the weaker constraint $\Omega h^2 < 0.2$, while the dashed lines impose $0.05 <\Omega h^2 < 0.15$. }
\label{fig:efficiency}
\end{figure}

Finally, we can now compare the efficiency of the different types of scans. For this purpose we use two definitions of 'good' points: (i) $\Omega h^2 < 0.2$ and (ii) $0.05 <\Omega h^2 < 0.15$. The results are summarised in \cref{fig:efficiency}. One can see that the MLS approach can reach efficiencies up to 70\% which is much better than a random or grid scan, with efficiencies of at most a few percent.

\section{Summary and Remarks}
\label{sec:summary}
A brief introduction to the first version of the scanning tool \xBIT for BSM models has been given here. The main motivation for \xBIT was to create a lightweight  tool which is easy to install and simple to use. Nevertheless, it is supposed to
provide a functionality which is already sufficient for many users in order to explore the parameter space of a (new) BSM model. \xBIT supports simple grid as well as random scans. Moreover, the MLS approach, which is an adaptive sampling algorithm based on artificial neural networks, has been implemented. \\[5mm]

{\it Remark}: \xBIT is still very young but I decided to publish it now because I'm going to stop developing HEP packages. This version of \xBIT can already be taken as replacement for the {\tt Mathematica} package {\tt SSP} which was used in the past by a number of people. Moreover, it might serve as basis for further developments. It is fully open source and written in a modular way which makes it possible to include new classes. Thus, I happily invite people to contribute new functionality if they think it's necessary -- and to share it with the community! Some extensions could be the link to Monte-Carlo and recasting tools, including other public libraries such as {\tt pyMultiNest}, or developing other machine learning approaches to sample parameter spaces. I think, reinforcement lerarning could be a promising direction. Have fun! \\[5mm]

{\it Note}: This package is completely independent of another sampling tool which comes with a bunch of 'BITs'. Nevertheless, this name was chosen because it might be a good complement since it follows in many respects a philosophy which is orthogonal to this other tool.

\section*{Acknowledgements}
I'm grateful to Jie Ren for helpful discussions about the MLS approach and technical advises. 
This work is supported by the ERC Recognition Award ERC-RA-0008 of the Helmholtz Association.

\begin{appendix}
\section{Available types of Scans}
\label{sec:scans}

\subsection{Random Scan}
\label{sec:random}
The simplest possibility for a scan is to vary all parameters independently and randomly in a given scan range. In order to generate the random numbers, one can make use of the different routines available in the different \python packages. The current implementation of the random scan assume that {\tt numpy} as used (and imported as {\tt np}). Thus, for getting uniformly distributed random numbers, one can use 
\begin{lstlisting}[language=ipython]
np.random.uniform(min, max) 
\end{lstlisting}
{\tt  numpy} offers many other distributed (normal, Poisson, binomial,\dots ). An dedicated function for a uniformly logarithmic distribution doesn't exist, but one can use  instead 
\begin{lstlisting}[language=ipython]
np.exp(np.random.uniform(np.log(min),np.log(max)))
\end{lstlisting}

\subsection{Grid Scan}
\label{sec:grid}
Another simple scan technique usually used for parameter spaces with a low dimensionality are grid scans. In that case, one needs to provide lists with the values the shall be considered for each parameter. \xBIT generates then all possible combinations. One can either provide a list including the values which shall be used, e.g. via 
\begin{lstlisting}[language=ipython]
[n1, n2, n3, ..., nN] 
\end{lstlisting}
or one can use one of the many possibilities to generate such lists based on a given distribution. Again, the default setting is to use {\tt numpy}. A linear distributed list of $N$ entries between two values is generated by
\begin{lstlisting}[language=ipython]
np.linspace(min, max, num=N) 
\end{lstlisting}
and for a logarithmic distribution one can use 
\begin{lstlisting}[language=ipython]
np.geomspace(min, max, num=N) 
\end{lstlisting}
Don't confuse {\tt geomspace} with {\tt logspace} which also exists in {\tt numpy}. Both routines can give the same result, but the input is different. For instance, the two inputs to get $\{10,100,1000,10000\}$ are
\begin{lstlisting}[language=ipython]
np.geomspace(10, 10000, num=4)
np.logspace(1, 4, num=4) 
\end{lstlisting}
Of course, also all other distributions available in {\tt numpy} can be used as well.

\subsection{Machine Learning Scan (MLS)}
\label{sec:MLS}
This scan is based on the idea presented in Ref.\cite{Ren:2017ymm}. The main purpose is to increase the 'efficiency' of a scan. Here, 'efficiency' means the ratio of 'good' points over all calculated points. 
'Good' points are those with a large likelihood with respect to some defined properties. MLS is an iterative approach which can be summarised as follows:
\begin{enumerate}
 \item An initial sample of points is generated using a random scan
 \item A neural network is trained with the data sample
 \item The sample of points is increased with points proposed by the neural network (90\%), while  only 10\% of the new points are chosen randomly
\end{enumerate}
The second and third step are iterated until the demanded number of points has been sampled. The simplest version of the MLS approach uses one deep, fully connected neural network for regression ('predictor'). 
In \xBIT a second neural network can be included which distinguishes valid and invalid points ('classifier'). Invalid points are those which either don't create any spectrum at all or for which the calculation of at least one observable fails. Moreover, it turned out that a mechanism is sometimes helpful to prevent the oversampling of parameter regions. In order to force the neural network to consider regions with a slightly lower likelihood, a penalty $\Delta$ can be included which results in 
an effective likelihood $\mathcal{L}_{\rm eff}$:
\begin{equation}
\mathcal{L}_{\rm eff} = \Delta \times \mathcal{L}_{\rm obs}  
\end{equation}
Here, $\mathcal{L}_{\rm obs}$ is the likelihood with respect to the defined variables and $\Delta$ is calculated as
\begin{equation}
\Delta(x) = \text{max}\left|\frac{ \text{min}(x_i - y_{i})}{\delta_i} \right|^2
\end{equation}
Here, $x$ is the point currently considered by the neural network (a vector in parameter space), $x_i$ is the $i$th component of this point, $y_i$ is the $i$th component of an already sampled point, and $\delta_i$ is the size of the scan range in this $i$th dimension. \\
\xBIT allows to set up the topology of the neural network, i.e. the number of neurons in the hidden layers, via the input file, see \cref{sec:setup}. Moreover, the learning rate of the Adam optimiser as well as maximal number of epochs can be set in the input. Out of the box, more details as the dropout (fixed to 10\%), and early stopping are defined in the file {\tt ml.py}. If necessary, these settings can be changed by modifying the class {\tt NN}.

\subsection{Marcov Chain Monte Carlo (MCMC)}
\label{sec:MCMC}
\xBIT also includes a basic implementation of the Metropolis–Hastings algorithm: 
\begin{enumerate}
 \item A random point is used for initialisation
 \item In the neighbourhood of the currently considered point a new point is tested
 \item The chain 'jumps' to the new point if 
 \begin{equation}
  \mathcal{L}_{\rm new} >  \mathcal{L}_{\rm old}
 \end{equation}
 or 
 \begin{equation}
  \frac{\mathcal{L}_{\rm new}}{\mathcal{L}_{\rm old}} > u
 \end{equation}
 where $n$ is a random number in the interval [0,1].
\end{enumerate}
The steps (2) and (3) are iterated until a criterion for abortion is satisfied. Right now, \xBIT just counts the number of points and stops after a given number of points has been sampled. Nevertheless, this basic approach is often successful to find interesting parameter regions even in models with a high-dimensional parameter space.

\section{Defining new classes}
\label{app:classes}
\subsection{Definition of a tools class}
\label{app:tools}
In order to implement the support of a new tool in \xBIT, it is sufficient to add a \python file into {\tt package/tools} which includes the commands to run that tool and to digest the results. For instance, the implementation of \HB looks as follows:
\begin{lstlisting}[language=ipython]
import os
import package.debug  as debug

class NewTool():
    def __init__(self):
        self.name = "HiggsBounds" 

    def run(self, settings, spc_file, temp_dir, log):
        # Settings
        command = settings['Command']
        options = settings['Options']
        output_file = settings['OutputFile']

        # Clean up
        if os.path.exists(output_file):
            os.remove(output_file)

        # Run HiggsBounds
        debug.command_line_log(command + " " + options + " " 
				+ temp_dir + "/", log)

        # Reading the Output file
        if os.path.exists(output_file):
            for line in open(output_file):
                li = line.strip()
                if not li.startswith("#"):
                    results = list(filter(None, line.rstrip().split(' ')))
            # Append output to the SPheno file
            debug.command_line_log("echo \"Block " + self.name.upper() + " # \" 
			      >> " + spc_file, log)
            for i in range(1, len(results)):
                debug.command_line_log("echo \"" + str(i) + " " + str(results[i])
                                    + " # \" >> " + spc_file, log)
        else:
            log.error("HiggsBounds output not written!",
                    command + " " + options + " " + temp_dir)
\end{lstlisting}
We see here the main ingredients necessary to set up a new tool:
\begin{enumerate}
 \item Define the class {\tt NewTool}
 \item Define the variable {\tt self.name} in the init function of this class. 
 \item Define the {\tt run} method. This method must include all commands to run the tool and to handle the output. The input parameters of this method are:
  \begin{itemize}
   \item {\tt settings}: these are the entries given in the settings file in the block named after the tool, see \cref{sec:settings}
   \item {\tt spc\_file}: the spectrum file which contains all necessary input parameters and which must be extended by the results of the given tool 
   \item {\tt temp\_dir}: the temporary directory in which the current scan is performed 
   \item {\tt log}: the logger object responsible for writng the log files
  \end{itemize}
\end{enumerate}
We see at this example that executing the tool is usually very simple. One can use for that the function {\tt debug.command\_line\_log} with two arguments: (i) the terminal command to run the tool, (ii) the logger object. Under the hood {\tt debug.command\_line\_log} is a wrapper function for {\tt subprocess.call} which pipes the output to the correct log file. The bigger part of this routine is responsible for reading the \HB results and for attaching them to the spectrum file in a SLHA-like format. 

\subsection{Definition of a scan class}
\label{app:scans}
In order to define a new type of scan, the necessary \python code needs to be put in {\tt package/scans}. For instance, the definition of the grid scan in \xBIT reads
\begin{lstlisting}[language=ipython]
from package.scanning import Scan as Scan
import itertools
import numpy

scan_name = "Grid"

class NewScan(Scan):
    """Scanner class for Grid Scans"""

    def __init__(self, inputs, config):
        Scan.__init__(self, inputs, config)

    def run(self):
        self.runner.run(self.config.log, sample=[])

    def generate_parameter_points(self, vars, points):
        all_points = []
        for x in vars:
            all_points.append(eval(vars[x]))
        
        temp = list(itertools.product(*all_points))
        all_points = [list(xx) for xx in temp]
        self.inputs['Setup']['Points'] = len(all_points)
        
        return all_points 
\end{lstlisting}
The main ingredients to a define a new kind of scan are:
\begin{enumerate}
 \item The variable {\tt scan\_name}. Note, this variable is not part of the class because one needs to have access before an instantiation of the class.
 \item The class {\tt NewScan}. This class is based on the class {\tt Scan} defined in {\tt package/scanning.py}. The input parameters are 
 \begin{itemize}
  \item {\tt inputs}: the content of the scan file, see \cref{sec:scanfile}
  \item {\tt config}: the reference to the configuration class which includes the reference to the logger and steers the screen output for instance
 \end{itemize}
 The base class {\tt Scan} provides already routines for necessary tasks as parsing the input file,
 creating the output directory, etc. There shouldn't be any need to modify these routines. However, two class methods must be given:
 \begin{itemize}
 \item {\tt run()}: as the name suggests, this is the routine to run the actual scan. In the given example, this just calls another routine {\tt self.runner.run}. This routine runs the full sample of points for all defined codes. (Take a look at the {\tt Runner} class in {\tt running.py} to see the details). For other scans, the {\tt run} method might look much more complicated and could for instance involve several iterations of the calls of {\tt self.runner.run} and training after each iterations. That's for instance done in the MLS scan. 
 \item {\tt generate\_parameter\_points()}: this routine returns the parameter points for a given scan. As input it takes:
 \begin{itemize}
  \item {\tt vars}: the {\tt Variables} block in the input file, see \cref{sec:scanfile}.
  \item {\tt points}: the number of points
 \end{itemize}
 Note, for a grid scan the variables {\tt points} isn't used at all, but the number of points is actually an output  (\verb"self.inputs['Setup']['Points'] = len(all)"). The rest of this routine evaluates the functions given in the input file for the different variables. Afterwards all possible combinations are generated by using the function {\tt product} of the {\tt itertools} package. 
\end{itemize}
\end{enumerate}
For a random scan, the function {\tt generate\_parameter\_points} is very simple:
\begin{lstlisting}[language=ipython]
    def generate_parameter_points(self, vars, points):
        all_points = []
        for _ in range(points):
            all_points += [[eval(vars[xx]) for xx in vars]]
        return all_points
\end{lstlisting}

\end{appendix}

\bibliographystyle{ArXiv}
\bibliography{lit}

\end{document}